\documentclass[aps,pra,preprint,groupedaddress,showpacs]{revtex4}
\begin{document}

\title{A geometric approach to the canonical reformulation of quantum mechanics}

\author{Mohammad Mehrafarin}
\affiliation{Physics Department, Amirkabir University of Technology, Tehran 15914,
Iran}
\email{mehrafar@aut.ac.ir}

\date{\today}

\begin{abstract}
The measure of distinguishability between two neighboring preparations of a physical system by a measurement apparatus naturally defines the line element of the preparation space of the system. We point out that quantum mechanics can be derived from the invariance of this line element in the canonical formulation. The canonical formulation of quantum statistical mechanics is also discussed.
\end{abstract}

\pacs{03.65.Ca, 03.65.Ta}

\maketitle

\section{Introduction}

Quantum mechanical descriptions refer to the complementary contexts set by incompatible measurements and it is with respect to such a context that preparations (viz. prepared states of a physical system) are described. This contextuality of description is manifest in the standard formulation of quantum mechanics: Different bases (representations) for the Hilbert space are provided by incompatible measurements through the eigenstates $\{|i>\}$ of the measured observable; where $i=1,...,n$ labels the measurement results; and a change of basis (representation), thus, corresponds to a change of measurement context. A state vector $|\psi>$, which represents a preparation, is described with respect to such a basis by complex components $\psi_i= \sqrt{p_i}\ e^{i \phi_i}$; where $\{p_i\}$ denotes the probability distribution of the measurement results and $\phi_i$ is the quantum phase of the preparation corresponding to the result $i$. The collection of (real) values $(p_i,\phi_i)$, therefore, describes an arbitrary preparation with respect to a given measurement context (measurement apparatus). The totality of all preparations may be called the preparation space of the system and an arbitrary preparation can, hence, be represented by a point in this space with coordinates $(p_i,\phi_i)$ relative to a given measurement context. Measurement apparatus (contexts), thus, provide reference frames or coordinate systems for the preparation space relative to which a point (preparation) is described by the probability distribution of the measurement results and the corresponding phases as its coordinates. Note, however, that, in spite of the contextuality of quantum mechanical descriptions, every reference frame (measurement context) is equivalent to every other frame (context) with regard to the description; there is no preferred measurement.

Now, there is a natural measure of distinguishability between two state vectors (preparations) in the Hilbert space, namely, the angle between the corresponding rays. For two neighboring preparations, 
$$[\cos^{-1}  |<\psi|\psi+d \psi>|]^2= \sum_{i} \frac{dp_i^2}{4p_i}+ \sum_{i} p_i d\phi_i^2-(\sum_{i} p_i d\phi_i)^2+higher\ order\ terms. $$
Since any two preparations in the preparation space are naturally discriminated by their distance, the above angle defines the (Riemannian) line element of the preparation space according to,  
\begin{equation}
ds^2=\sum_{i} \frac{dp_i^2}{4p_i}+\sum_{i} p_i d\phi_i^2-(\sum_{i} p_i d\phi_i)^2.
\end{equation}
This line element (also known as the Fubini-Study metric \cite{Anandan,Anandan2,Gibbons,Braunstein}) represents the measure of distinguishability by a given measurement apparatus between two neighboring preparations $(p_i,\phi_i)$ and $(p_i+dp_i, \phi_i+d\phi_i)$. However, because all measurement contexts are equivalent with regard to the description of a given preparation and there is no preferred measurement, the measure of distinguishability has to be invariant with respect to all measurement apparatus. (Otherwise, some measurements would be more discriminating than others would, which provide a basis for preference.)

In the framework provided by the preparation space, we point out in section II that quantum mechanics can be derived from the invariance of the line element in the canonical formulation. The first term in the line element is the well-known measure of distinguishability between two neighboring probability distributions \cite{Fisher, Bhattacharyya, Wootters} and the second term is the variance of the phase difference. Although the canonical formulation is well known \cite{Ashtekar,Weinberg,Hall,Hall2,Landsman,Guerra,Minic,Minic2,Minic3}, our emphasis on the fundamental significance of the line element bears valuable insight for understanding the foundations of quantum mechanics, provided the line element can be obtained from an independent premise. Such a premise, from which the line element follows as the sum of the above two terms, is given in reference \cite{Mehrafarin}. Finally, in section III we discuss the canonical formulation of quantum statistical mechanics in the preparation space.

\section{Canonical quantum mechanics}
Here we briefly review the elements of canonical quantum theory with emphasis on the role of invariance of line element (1). The invariance of the line element restricts the form of the allowed coordinate transformations $(p_i,\phi_i) \rightarrow (p_i^\prime,\phi_i^\prime)$ in the preparation space. Such transformations will determine how a given preparation is to be described with respect to different measurement contexts. We do not need to work out the transformation law from scratch. It was shown by Wigner \cite{Wigner} that the most general angle-preserving transformation in the Hilbert space is a unitary transformation. Hence, the required transformations in the preparation space correspond to the unitary transformations of the standard formulation associated with a change of basis. Writing the unitary transformation matrix as $u_{ji}= \sqrt {\omega_{ij}}\ e^{i \beta_{ij}},$ the unitary transformation $\psi_i^\prime= \sum_{j} u_{ji}^* \psi_j$ reads:
\begin{eqnarray}
p_i^\prime= \sum_{jk} \sqrt{\omega_{ij}p_j}\ \sqrt{\omega_{ik}p_k}\ \cos (\phi_{jk}-\beta_{ij}+\beta_{ik}), \nonumber \\
\tan \phi_i^\prime= \frac {\sum_{j} \sqrt{\omega_{ij}p_j}\ \sin (\phi_j-\beta_{ij})}{\sum_{j} \sqrt{\omega_{ij}p_j}\ \cos (\phi_j-\beta_{ij})},\ \ \ \ \ \ 
\end{eqnarray}
where $\phi_{jk}=\phi_j-\phi_k$ is a relative phase, and the $2n^2$ transformation parameters $\omega_{ij}$ and $\beta_{ij}$ satisfy,
\begin{eqnarray}
\sum_{i} \sqrt {\omega_{ij}\omega_{ik}}\ \matrix {\cos\cr \sin\cr}(\beta_{ik}-\beta_{ij})&=& \sum_{i} \sqrt {\omega_{ji}\omega_{ki}}\ \matrix {\cos\cr \sin\cr} (\beta_{ki}-\beta_{ji})=0, \ \ (j \neq k) \nonumber \\
\sum_{i} &\omega_{ij}&= \sum_{i} \omega_{ji}=1. 
\end{eqnarray}
These are $n^2$ constraints, which translate the unitary conditions
$\sum_{i} u_{ij}^* u_{ik}= \sum_{i} u_{ji} u_{ki}^*= \delta_{jk},$ leaving only $n^2$ transformation parameters independent. 

Coordinate transformation (2) is the required transformation law in the preparation space, which relates, in terms of $n^2$ independent parameters, the descriptions of a given preparation with respect to different measurement contexts. We next show that the evolution law of an isolated preparation (the Shr\"{o}dinger equation) naturally follows, too, through the same invariance property.

In the preparation space of an isolated system, consider an arbitrary preparation specified by the coordinates $(p_i,\phi_i)$ with respect to a given measurement apparatus. Because there is no preferred frame, the equations governing the time development of the preparation have to be covariant with respect to the transformation law (2). Now from (2), it follows after some calculations using conditions (3), that,
\begin{equation}
MJM^T=J,
\end{equation}
where the $(2n \times 2n)$ matrices $M$ and $J$ are given by,
\begin{equation}
M=\left( \matrix{\partial p_i^\prime/\partial p_j &\partial  p_i^\prime/\partial \phi_j \cr
\partial \phi_i^\prime/\partial p_j &\partial \phi_i^\prime/\partial \phi_j 
\cr} \right), \ \ \ \ \ J=\left( \matrix{0  &\delta_{ij}\cr   
-\delta_{ij}  &0\cr} \right).
\end{equation}
Equation (4) is recognized as expressing the necessary and sufficient condition (the symplectic condition \cite{Goldstein}) for the canonicality of the coordinate transformation (2); i.e.; the necessary and sufficient condition for the covariance of the Hamilton-like equations,
\begin{equation}
\dot{p_i}= \frac {\partial {\cal H}}{\partial \phi_i}, \ \ \ \ \dot{\phi_i}= -   \frac {\partial {\cal H}}{\partial p_i},
\end{equation}
under the transformation; ${\cal H}$ being a scalar
(${\cal H} (p_i,\phi_i,t)= {\cal H^\prime} (p_i^\prime,\phi_i^\prime,t)$)
with the dimensions of $time^{-1}$, of course. Adopting units $\hbar=1$, clearly,  the Hamiltonian ${\cal H}$ should be identified with the mean (expectation) energy of the preparation which has the same value in all frames (representations). The canonical equations (6), being the only covariant set of equations under (2), then provide a unique candidate for the `equations of motion' of the preparation. Making contact with the standard formulation of quantum mechanics, we have,
$$ {\cal H} (p_i,\phi_i,t)=<\psi|H|\psi>= \sum_{ij} H_{ij}(t) \sqrt{p_i p_j}\ e^{-i \phi_{ij}},$$
where $H$ is the Hamiltonian operator (whose possible time dependence results in an explicit time dependence of ${\cal H}$). Whence, (6) translates into
$i \dot{\psi_i}= \sum_{j} H_{ij} \psi_j,$
which is just the Shr\"{o}dinger equation in the representation provided by the measurement apparatus; the covariance of (6) under the transformation (2) corresponds to the covariance of the latter under unitary transformations. 

Needless to emphasize, the unitary transformation group of quantum mechanics on the one hand, and the Shr\"{o}dinger equation on the other, both have emerged through the invariance property of the preparation space. 

The dynamics in the preparation space, whence, closely resembles classical dynamics in the phase space picture. In particular, the canonically conjugate coordinates $(p_i,\phi_i)$ determine the evolution trajectory of a preparation in the preparation space of an isolated system with mean energy ${\cal H}$. Furthermore, due to the time development of the preparation, the mean value of an arbitrary observable $F$, namely the scalar,
$$f(p_i,\phi_i,t) \equiv <\psi|F|\psi>=\sum_{ij} F_{ij}(t) \sqrt{p_i p_j}\ e^{-i \phi_{ij}},$$
thus becomes a dynamical variable in the preparation space. Its dynamics follows from the equations of motion (6) to be determined from,
\begin{equation}
\dot{f}=\frac {\partial f}{\partial t}+\sum_{i} (\frac {\partial f}{\partial p_i} \frac {\partial {\cal H}}{\partial \phi_i}-\frac {\partial f}{\partial \phi_i} \frac {\partial {\cal H}}{\partial p_i}) \equiv \frac {\partial f}{\partial t}+\{f,{\cal H}\},
\end{equation}
where $\{f,{\cal H}\}$ denotes the Poisson bracket of $f$ and ${\cal H}$. Needless to say, because Poisson brackets are invariant under canonical transformations, the dynamics is independent of the choice of the reference frame of the measurement apparatus. Equation (7), of course, corresponds to the equation
$\dot{f}=<\dot{F}>+\frac{1}{i} <[F,H]>$
of the standard formulation, since,
$$<\dot{F}>=\sum_{ij} \dot{F_{ij}} \sqrt{p_i p_j}\ e^{-i \phi_{ij}}=\frac {\partial f}{\partial t},$$
and
\begin{equation}
\frac{1}{i} <[F,H]>=\{<F>,<H>\}=\{f,{\cal H}\},
\end{equation}
as can be demonstrated directly.

\section{Canonical quantum statistical mechanics}
A point in the preparation space corresponds to a pure state, i.e. to maximal information about the system at a given time compatible with objective data from the measurements of all conceivable experiments on the observables of the system. Ideally, when such a prior information is available as initial data, the subsequent time evolution of the state is determined by the Hamilton-like equations of motion. Otherwise, the time development of physical quantities of interest has to be inferred from incomplete information. We will show that the canonical approach leads to a reformulation of quantum statistical mechanics that resembles the phase space formulation of classical statistical mechanics. 

Taking advantage of the canonical formulation, in analogy with classical mechanics in phase space, the time evolution of an isolated preparation can be represented by a succession of infinitesimal canonical transformations generated by ${\cal H}$. Then, because under canonical transformations, 
$$\delta(\sum_{i} p^\prime_i-1) d^n p^\prime\ d^n \phi^\prime=\delta(\sum_{i} p_i-1)\ \|M\| d^n p\ d^n \phi=\delta(\sum_{i} p_i-1) d^n p\ d^n \phi,$$
the volume element, $d\mu=\delta(\sum_{i} p_i-1) d^n p\ d^n \phi$, of the preparation space remains invariant in time. Thence, it follows that the probability distribution of points, $w(p_i,\phi_i,t)$, in the preparation space is also a constant of motion, i.e.,
\begin{equation}
\dot{w}=\frac {\partial w}{\partial t}+\{w,{\cal H}\}=0.
\end{equation}
This Liouville-like equation is relevant when maximal information is not available to determine the preparation uniquely and one, therefore, must deal with a probability distribution of preparations (a mixed ensemble) compatible with the information given. Suppose the prior information is composed of the experimental result of the measurement of some observable $F$ of the system, which is expressed in terms of its expectation value $\bar{F}$ at $t=0$ in the standard manner (the generalization to a number of commuting observables is straightforward). The information entropy associated with this measurement is,
$$S=-\sum_{i=1}^{n} \rho_i \ln \rho_i,$$
where $\rho_i$ is the probability of obtaining the eigenvalue $F_i$ of $F$. The probabilities $\{\rho_i\}$ that best describe the information are obtained by maximizing the entropy subject to the measurement result,
\begin{equation}
\sum_{i} \rho_i F_i=\bar{F},
\end{equation}
and the normalization condition,
\begin{equation}
\sum_{i} \rho_i =1.
\end{equation}
The result is,
\begin{equation}
\rho_i =\frac{e^{-\beta F_i}}{Z},
\end{equation}
where $Z=\sum_{i} e^{-\beta F_i}$ is the partition function and $\beta$ is determined by condition (10), which may now be written as,
\begin{equation}
-\frac{\partial}{\partial\beta} \ln Z=\bar{F}.
\end{equation}
Now, with respect to the measurement context under consideration, an arbitrary preparation in the ensemble has coordinates $(p_i,\phi_i)$, where $p_i$ is the probability associated with the result $F_i$. Hence, weighing by the corresponding probabilities $w$ of the preparations we arrive at,
\begin{equation}
\rho_i=\int wp_i\ d\mu,
\end{equation}
the integral being over all space ($0\leq p_i \leq 1, 0\leq \phi_i \leq 2\pi$). Constraints (10) and (11), thus, translate into constraints on $w$ according to,
\begin{eqnarray}
\int wf\ d\mu=\bar{F},\\
\int w\ d\mu=1,\nonumber \ \
\end{eqnarray}
respectively ($f=<F>=\sum_i p_i F_i$). The canonical invariance of these is manifest ($w$, $f$ and $d\mu$ are all scalar). With $\rho_i$ given by (12), the integral equation (14) can be solved for $w$ to yield,
\begin{equation}
w_0(p_i,\phi_i)=(n+1) \frac {<e^{-\beta F}>}{\int <e^{-\beta F}> d\mu}-\frac {n!}{(2\pi)^n}.
\end{equation}
Condition (15), which determines $\beta$, may now be written as,
$$-\frac{\partial}{\partial\beta} \ln \int <e^{-\beta F}> d\mu=\bar{F}.$$
This is of course just equation (13), since,
$$\int <e^{-\beta F}> d\mu=\sum_i e^{-\beta F_i} \int p_i d\mu= \frac{(2\pi)^n}{n!}\ Z.$$
(In the above manipulations the result,
$$\int_0^1 p_1^{m_1}. . .\ p_n^{m_n}\ \delta(\sum_i p_i -1)\ d^n p=\frac {\prod_i (m_i!)}{(\sum_i m_i+n-1)!},$$
is useful.) Distribution (16) in the preparation space represents the measurement result at $t=0$. It can be used as initial value for the Liouville-like equation (9) to yield $w(p_i,\phi_i,t)$. Then, the expectation value of any observable $Q$ at arbitrary time $t$ will be given by,
$$\int w(p_i,\phi_i,t) q\ d\mu=\bar{Q}(t),$$
where, of course, $q(p_i,\phi_i)=<Q>$ is the corresponding dynamical variable in the preparation space. For equilibrium distributions, $\partial_t w=0$, so that by equation (9), $\{w,{\cal H}\}=0$. Since, from (8),
$$\{<e^{-\beta F}>,{\cal H}\}=\frac{1}{i} <[e^{-\beta F},H]>,$$
it follows that if $F$ is a constant of motion, the distribution given by (16) will be an equilibrium distribution. An immediate example is provided by the canonical distribution for which $F=H$.

It is instructive now to make direct correspondence with the standard formulation in terms of the density operator, $\rho$. By definition, with respect to an arbitrary measurement frame,
\begin{equation}
\rho_{ij}(t)=\int w\psi_i \psi_j^*\ d\mu= \int w(p_i,\phi_i,t) \sqrt{p_i p_j}\ e^{i \phi_{ij}}\ d\mu.
\end{equation}
It follows that,
\begin{eqnarray}
tr \rho=\sum_i \rho_{ii}=\int w\ d\mu,\ \ \ \ \ \nonumber\\
tr(\rho F)=\sum_{ij} \rho_{ij} F_{ji}=\int wf\ d\mu. \nonumber
\end{eqnarray}
The normalization condition and the expectation formula, thus, reduce to their familiar expressions in the standard formulation. Furthermore, if the reference frame is such that phases are absent in the functional form of $w$, then, upon performing the phase integration, (17) reduces to $\rho_{ij}=\rho_i \delta_{ij}$, where $\rho_i$ is given by (14). This corresponds to the diagonal representation of $\rho$, with $\{\rho_i\}$ as its eigenvalues. Therefore, the entropy too reduces to its familiar expression $S=-tr(\rho\ ln \rho)$. Finally, to complete the correspondence we should prove the equivalence of the von Neumann equation for $\rho_{ij}$ and the Liouville-like equation for $w$. One approach would be to substitute (17) into the von Neumann equation and obtain the Liouville-like equation (9) as the necessary and sufficient condition. However, a simpler proof is provided by noting that, since,
$$(2\pi)^n w=(n+1)! <\rho>-n!,$$
which is most easily derived from (17) in the diagonal representation, we have
$$\dot{w}=\frac{(n+1)!}{(2\pi)^n}<\dot{\rho}+\frac{1}{i}\ [\rho,H]>.$$
Hence, von Neumann equation implies and is implied by $\dot{w}=0$, because the average is over arbitrary preparation.

In the canonical quantum statistical mechanics, the probability distribution of preparations and its Liouville-like equation replace the density operator and the von Neumann equation of the standard formulation, as we have shown. The canonical reformulation closely resembles classical statistical mechanics apart from the expression for the entropy, namely,
$$S=-\int w <ln \rho>\ d\mu,$$
due to the existence of quantum probabilities $p_i$ ($S_{class}=-\int w\ lnw\ d\mu$).

\end{document}